\documentclass[12pt]{article}
\usepackage{a4wide,graphicx,amsmath,amssymb,hyperref}
\usepackage[numbers,sort&compress]{natbib}
\usepackage{hypernat}
\usepackage{subfigure}
\usepackage{multirow}
\usepackage{axodraw}
\def\slash#1{#1 \hskip-0.45em /}
\def\beq{\begin{equation}}
\def\eeq{\end{equation}}
\def\bea{\begin{eqnarray}}
\def\eea{\end{eqnarray}}

\newcommand{\newc}{\newcommand}
\newc{\MW}{M_W}

\newc{\CKMp}{\tilde{V}}
\newc{\CKM}{V_{\rm CKM}}
\newc{\cp}{\tilde{\theta}_c}
\newc{\ifb}{\textrm{fb}^{-1}}
\newc{\fb}{\textrm{fb}}
\newc{\Wprime}{W'}
\newc{\wmp}{M_{W'}}
\newc{\Wp}{W^+}
\newc{\Wm}{W^-}
\newc{\Wpm}{W^{\pm}}
\newc{\Maj}{N_l}
\newc{\nm}{M_{N}}

\newc{\ttbar}{t\bar{t}}
\newc{\Wbbj}{Wb\bar{b}j}
\newc{\WZjj}{WZ(\gamma^*)jj}
\newc{\kkbar}{K^0-\bar{K}^0}
\newc{\Zgam}{Z(\gamma^*)}

\newc{\nubb}{0\nu\beta\beta}

\newc{\ord}{{\mathcal O}}

\newc{\Rpm}{R_{\pm}}
\newc{\pRpm}{\tilde{R}_{\pm}}
\newc{\Np}{N_+}
\newc{\Nm}{N_-}

\newc{\Lag}{{\mathcal L}}

\newc{\HT}{H_T}
\newc{\ptl}{p^l_T}
\newc{\ptj}{p^j_T}
\newc{\mll}{m_{ll}}
\newc{\mlljj}{m_{lljj}}
\newc{\mwrec}{\wmp^{\rm rec}}
\newc{\MET}{\slash{E}_{T}}
\newc{\pt}{p_T}

\newc{\epsb}{\epsilon_{B}}

\newc{\herwig}{\texttt{Herwig++}}
\newc{\herwigv}{\texttt{Herwig++2.4.2}}
\newc{\fherwig}{\texttt{Herwig6.510}}
\newc{\fastjet}{\texttt{FastJet-2.4.2}}
\newc{\mcfm}{\texttt{MCFM}}
\newc{\mstwlo}{\texttt{MSTW08 LO}}
\newc{\vegas}{\texttt{VEGAS}}
\newc{\madgraph}{\texttt{MADGRAPH}}
\newc{\legacy}{\texttt{LEGACY}}
\newc{\alpgen}{\texttt{ALPGEN}}

\begin{document}
  \titlepage
  \begin{flushright}
    Cavendish-HEP-2010/18 \\
    October 2010 \\
  \end{flushright}
  \vspace*{0.5cm}
  \begin{center}
    {\Large \bf Charge asymmetry ratio as a probe of \\quark flavour
      couplings of resonant particles at the LHC}\\
    \vspace*{1cm}
    \textsc{C.H.~Kom and W.J.~Stirling}\\
    \vspace*{0.5cm}
         Cavendish Laboratory, University of Cambridge, CB3 0HE, UK\\
  \end{center}
  \vspace*{0.5cm}
  \begin{abstract}
    We show how a precise knowledge of parton distribution functions,
    in particular those of the $u$ and $d$ quarks, can be used to
    constrain a certain class of New Physics models in which new heavy
    charged resonances couple to quarks and leptons. We illustrate the
    method by considering a left-right symmetric model with a
    $\Wprime$ from a $SU(2)_R$ gauge sector produced in
    quark-antiquark annihilation and decaying into a charged lepton
    and a heavy Majorana neutrino. We discuss a number of quark and
    lepton mixing scenarios, and simulate both signals and backgrounds
    in order to determine the size of the expected charge
    asymmetry. We show that various quark--$\Wprime$ mixing scenarios
    can indeed be constrained by charge asymmetry measurements at the
    LHC, particularly at $\sqrt{s} = 14$~TeV.
  \end{abstract}


\section{Introduction}

Many theories beyond the Standard Model (SM) predict existence of new
particles that can be discovered at the LHC.  Once they are discovered
and basic properties such as masses measured, a next important step
would be to determine the couplings of these new particles.
Determination of these couplings will be crucial in unravelling the
structure of the underlying physics models.  While this is a
non-trivial task, strategies to measure various important
model-specific couplings have been proposed.

In this paper, we discuss how precise knowledge of parton distribution
functions (PDFs) could provide important information on a particular
class of couplings.  These are the couplings between quarks and
\emph{charged} resonances that might be produced at the LHC, i.e. 
via $q \bar q' \to X^{\pm}$.

The fact that the LHC is a proton-proton collider means that for many
processes, positively charged (leptonic) final states will be more
copiously produced than their negatively charged counterparts.
Depending on the typical scales involved, the charge asymmetry ratio
\bea 
\Rpm &\equiv& \frac{\Np}{\Nm}, 
\eea 
where $\Np$($\Nm$) are number of events with positive(negative) charge
assignment, for a given event topology can be determined if the
underlying hard process is known.\footnote{The paradigm process is of
  course Standard Model $W^\pm$ production, for which $\Rpm \sim
  1.33(1.42)$ at 14(7)~TeV LHC in NLO QCD perturbation
  theory~\cite{Kom:2010mv,Martin:2009iq}. Here the process is
  initiated primarily by $u\bar{d}$ and $d\bar{u}$ scattering.}  In
Ref.~\cite{Kom:2010mv} the variable $\Rpm$ was suggested as a
diagnostic tool for the presence of New Physics (NP) with a SM
background dominated by $W+n$ jet processes.\footnote{For
  state-of-the-art calculations of $W+n$ jet cross sections in the SM,
  the reader is referred to Refs.~\cite{MCFM,WnJ_NLODixon}.}  An
important advantage of using $\Rpm$ is that theoretical uncertainties
tend to cancel in the ratio, leading to relatively robust predictions.

Another feature of $\Rpm$ is that for (sub)processes produced at the
same scale, contributions from different incoming partons will
contribute to the ratio differently, due to the non-universal flavour
content inherent in the proton.  For example, processes initiated by
valence quarks $u$ and $d$ will tend to give $\Rpm > 1$, since $u(x) >
d(x)$, while those initiated by the sea quarks $c$, $s$ and $b$ will
lead to $\Rpm \sim 1$.  A measured value of $\Rpm$ can therefore lead
to a constraint on the relative contribution from these initial state
partons.  Moreover, as the ratio is independent of the absolute
normalisation of the couplings involved, it provides a genuine test of
the structure of the parton coupling to the final state
particle(s).\footnote{Of course the statistics and hence the accuracy
  in determing $\Rpm$ will still be affected by the normalisation of
  the couplings involved.}

These ideas can be applied to NP models where new charged particles
are produced at resonance at the LHC.  The presence of a resonance
implies that, in the absence of SM background, $\Rpm$ should track
closely the ratio of `weighted' parton luminosities
\beq\label{eq:partonLuminosity}
\frac{\partial \mathcal{L}_{ab}}{\partial M_V^2} = \frac{1}{s}
\int_{y_{\rm min}}^{y_{\rm max}}{\rm d}y\,|\CKMp_{ab}|^2f_a(x_1,M_V^2)f_b(x_2,M_V^2),
\eeq
where $x_1=\frac{M_V}{\sqrt{s}}e^y$, $x_2=\frac{M_V}{\sqrt{s}}e^{-y}$,
$M_V=\sqrt{x_1x_2s}$ is the mass of the resonance, $\CKMp_{ab}$ is its
coupling to partons $a$ and $b$, and $[y_{\rm min},y_{\rm max}]$ is
the rapidity interval in which the resonance is produced.  We denote
such parton luminosity ratio by $\pRpm$.  At higher orders, additional
QCD radiation will skew the parton luminosity prediction slightly, as
will cuts on the decay products of the resonance.  Nevertheless it
should still be possible to obtain quantitative estimates of $\Rpm$
using appropriate Monte Carlo programmes.

Charged resonances that might be discovered at the LHC include, for
example, vector bosons in technicolor or extended gauge theories
\cite{Ramond:1983qu,Altarelli:1989ff}, and charged sleptons in
standard supersymmetric models with R-parity violation
\cite{Dreiner:2000vf}.  In the same spirit, SM Higgs-stralung off a
(off-shell) $W$ boson is another interesting example, since we would
expect $\sigma(W^+H) > \sigma(W^-H)$.

For concreteness, we will focus in this study on a left-right (LR)
symmetric model \cite{LRSM} with a heavy $\Wprime$ from a $SU(2)_R$
gauge sector.  The primary leptonic decay mode of the $\Wprime$ is
$\Wprime \to l N_l$, where $\Maj$ is a heavy Majorana neutrino,
followed by three-body decay of $\Maj$ into a charged lepton and 2
jets.  The Majorana nature of $\Maj$ leads to `same-sign di-lepton
(SSDL) plus 2 jets' events.  This mode has not so far been searched
for in the Tevatron ($p \bar p$) experiments (however see the CDF
\cite{CDF_Wprime} and D0 \cite{D0_Wprime} $\Wprime$ searches in other
decay channels).  The discovery potential of this model at the 14~TeV
LHC has been studied in \cite{Aad:2009wy,Ferrari:2000sp}.  We will see
that the SM background can be strongly suppressed by applying cuts
similar to those in the above studies, together with the requirement
that the di-lepton pair has the same charge.  As a result, the
experimentally determined value of $\Rpm$ will be correlated with the
parton luminosity predictions.  Note that the charge asymmetry ratio
is only weakly dependent on the decay modes of the charged resonance.
Therefore our discussion can be readily extended to other $\Wprime$
models, for example the extended gauge theory model considered in
\cite{Ramond:1983qu,Altarelli:1989ff} (the discovery potential of this
model at the LHC is discussed in \cite{Aad:2009wy}), in addition to
the models mentioned above.

There exist models with additional constraining flavour structures,
for example discrete left-right symmetries relating the $SU(2)_L$ and
$SU(2)_R$ sectors.  Without a concrete high-scale model
implementation, the flavour/generation dependent couplings may be
considered arbitrary \cite{Amsler:2008zzb}.  Phenomenologically, if
$\Wprime$ is relatively light, the quark mixing structure can be
strongly constrained by low energy observables such as flavour
changing neutral currents (FCNC), for example from $\kkbar$ mixing
\cite{Beall:1981ze}.  We will illustrate to what extent different
flavour structures can be distinguished using $\Rpm$.

The paper is organised as follows. In the next section we specify our
$\Wprime$ model and discuss the behaviour of the parton luminosity
ratios $\pRpm$ given a set of representative quark flavour mixing
matrices.  In Section~\ref{sec:numerical} we then perform a numerical
study to show the relation between $\pRpm$ and $\Rpm$, followed by a
discussion on the extent to which $\Rpm$ can be used to distinguish
quark flavour structures at 14~TeV LHC.  Finally, we briefly comment
on the prospect of using this observable in the early stages of LHC
running.


\section{The $\Wprime$ model}
The Lagrangian describing the coupling of the $\Wprime$ to fermions
(quarks and leptons) is given by\footnote{A general interaction
  Lagrangian is given in \cite{Papaefstathiou:2009sr}.}

\bea\label{eq:WprimeLag}
\Lag_{\Wprime}&=&\frac{\tilde{g}}{\sqrt{2}}\CKMp_{ij}\bar{u}_{i}\gamma^{\mu}P_Rd_{j}W^{'+}_{\nu} +
\frac{\tilde{g}}{\sqrt{2}}\tilde{U}_{ij}\bar{e}_{i}\gamma^{\mu}P_RN_{j}W^{'-}_{\nu} + {\rm h.c.}\, \eea
In the above, we have assumed that the $\Wprime$ couples only to
right-handed fermions with an overall coupling strength normalised to
$\tilde{g}$.  The quark and lepton mixings are given by $\CKMp$ and
$\tilde{U}$ respectively.  As our focus is on the effect of the
structure of $\CKMp$ on $\Rpm$, we will assume, for simplicity, that
$\tilde{g}$ is the same as the SM electroweak coupling $g$ and that
$\tilde{U}$ is diagonal, i.e.  that the two charged leptons from
$\Wprime$ decay have the same flavour.  This allows us to focus on
same-sign, same-flavour leptonic final states.  However for more
general $\tilde{U}$, same-sign, \emph{different}-flavour final states
can also be considered.

In some (LR) symmetric models, $\CKMp$ is related to the SM CKM
matrix, $\CKM$.  In the limit of (pseudo-)manifest LR symmetry,
$\CKMp=\CKM$ and the mass of $\Wprime$ is constrained to be above a
few TeV \cite{Maiezza:2010ic}, primarily by $\Delta F=2$ processes
such as $\kkbar$ mixing.  However this lower bound can be relaxed by
considering more general $\CKMp$ structures, see for example
Ref.~\cite{Langacker:1989xa}.  A recent study \cite{Buras:2010pz}
suggests that if the quark mixing takes the form
\bea\label{eq:buras} |\CKMp| &\sim& \left( \begin{array}{ccc} 0 & \frac{1}{\sqrt{2}} &
  \frac{1}{\sqrt{2}} \\ 1 & 0 & 0 \\ 0 & \frac{1}{\sqrt{2}} &
  \frac{1}{\sqrt{2}} \end{array} \right)\,, \eea
then right-handed current contributions could lead to sizeable CP
violating effects in $B_s$ mixing as suggested by recent results
\cite{Aaltonen:2007he}, while remaining consistent with stringent
$\Delta F=2$ observables for relatively low effective scales.  We will
take this quark mixing structure, from now on denoted by $\CKMp_{\rm
  II}$, as one of our `reference' structures.  This can be compared
with other representative structures, for example CKM-like mixing
($\CKMp_{\rm CKM}$), diagonal mixing ($\CKMp_{\rm I}$) and when the
$\Wprime$ couples only to a particular generation ($\CKMp_{\rm
  1st}$,$\CKMp_{\rm 2nd}$).  These quark mixing patterns are listed in
Table~\ref{tab:mixingStruc}.\footnote{The matrices are normalised
  according to $\sum_{ij} |\CKMp_{ij}|^2 = 3$.}  We will compare to
what extent $\Rpm$ can distinguish these different mixing patterns.
\begin{table*}
  \centering
  \begin{tabular}{|c|c|}
    \hline
    & $\CKMp$ \\
    \hline\hline
    Diagonal mixing & $\CKMp_{\rm I}$ = $\mathbb{I}$\\
    CKM mixing & $\CKMp_{\rm CKM}$ = $\CKM$ \\
    Off-diagonal mixing & $\CKMp_{\rm II}$ (see eq.~(\ref{eq:buras}))\\
    1st generation only & $(\CKMp_{\rm 1st})_{ij}$ = $\sqrt{3}\delta_{i1}\delta_{j1}$\\
    2nd generation only & $(\CKMp_{\rm 2nd})_{ij}$ = $\sqrt{3}\delta_{i2}\delta_{j2}$\\
    \hline
  \end{tabular}\label{tab:mixingStruc}
  \caption{Representative quark mixing matrices $\CKMp$ in the charged
    current Lagrangian with a $\Wprime$, see
    Eq.~(\protect\ref{eq:WprimeLag}).  The normalisation of these
    matrices is such that $\sum_{ij} |\CKMp_{ij}|^2 = 3$.}
\end{table*}
For simplicity, we assume no mixing between the SM $W$ and the
$\Wprime$.  The decay mode $\Wprime \to WZ$ is forbidden as a result.

As discussed in the Introduction, once $\CKMp$ is specified, the ratio
of parton luminosities defined in Eq.~\ref{eq:partonLuminosity} can be
determined.  As long as the events selected to determine the charge
asymmetry ratio come dominantly from the $\Wprime$ signal, $\Rpm$
should track the corresponding $\pRpm$ closely.  We again note that
this correspondence should depend only weakly on the event topologies
and acceptance cuts.

\begin{figure*}[!ht]
  \begin{center}
    \scalebox{0.88}{\includegraphics{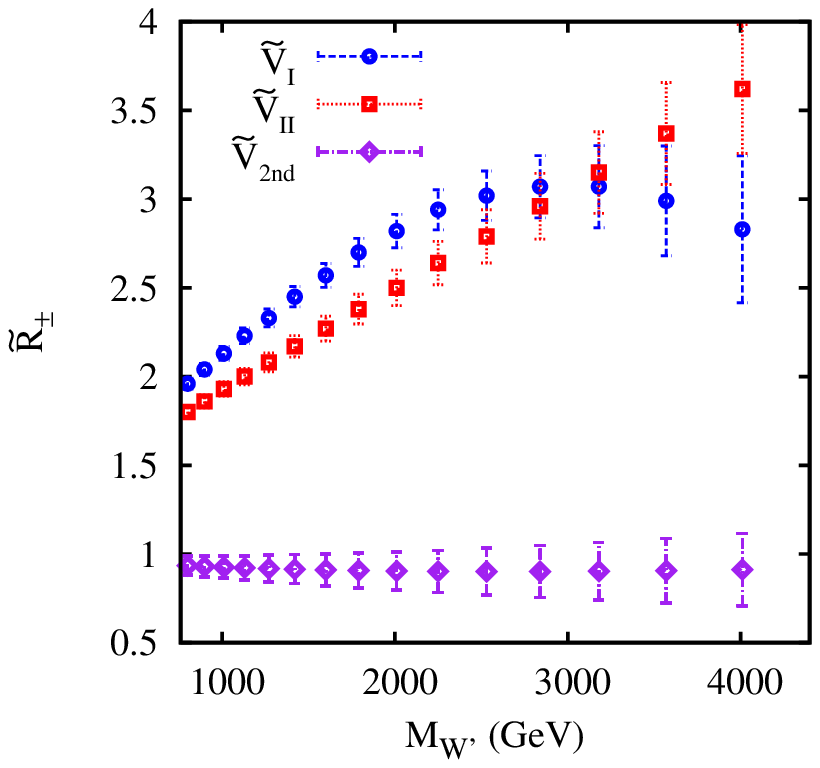}}
    \scalebox{0.88}{\includegraphics{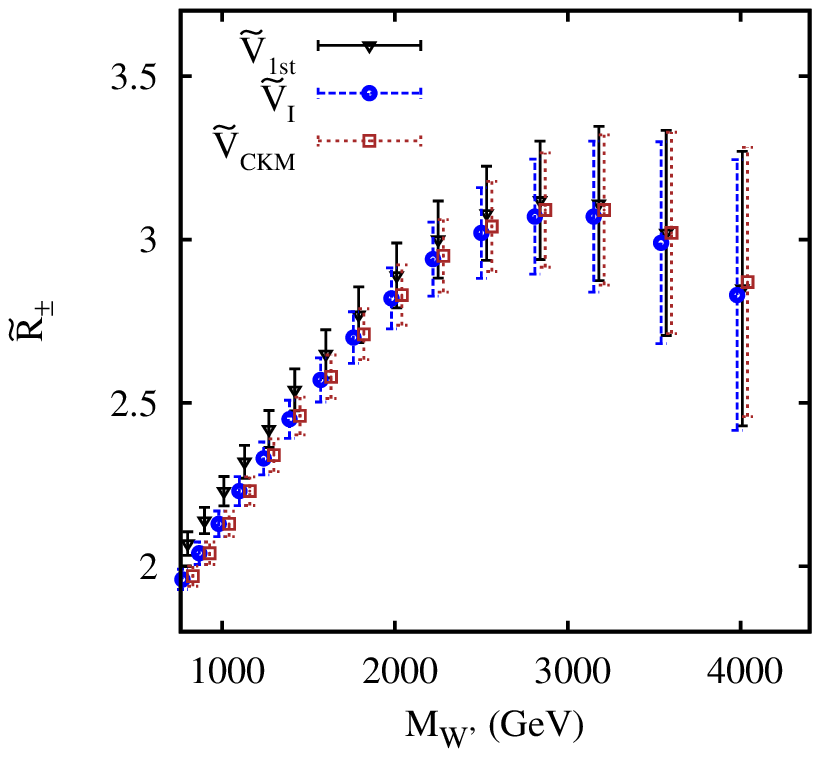}}
    \caption{Variation of parton luminosity ratios $\pRpm$ with $\wmp$
      at $14$~TeV LHC.  The quark mixing matrices $\CKMp$ are defined
      in Table~\protect\ref{tab:mixingStruc}.  The rapidity $y$ of
      $\Wprime$ satisfies $|y|<2.5$.  The ratio is computed using the MSTW08
      NLO PDF set \cite{Martin:2009iq} with factorisation scale $\mu_F = \wmp$.  The error
      bars correspond to 68\%cl PDF errors.}
    \label{fig:partonLum} 
  \end{center}
\end{figure*}

In Fig.~\ref{fig:partonLum} we display $\pRpm$ for the various quark
mixings shown in Table~\ref{tab:mixingStruc} as a function of $\wmp$.
In these plots, MSTW08 NLO PDF sets \cite{Martin:2009iq} are used,
with 68\%cl PDF uncertainties also included. Once NLO pQCD corrections
are accounted for, the PDF uncertainties are expected to be the most
important theoretical uncertainty on the ratio predictions. The
factorisation scale is chosen to be equal to the mass of the
resonance.  As we can see,the ratios for $\CKMp_{\rm 1st}$,
$\CKMp_{\rm I}$ and $\CKMp_{\rm CKM}$ are all quite similar, as they
are dominated by contributions from the first generation quarks.
However $\CKMp_{\rm II}$ shows a clear deviation from $\CKMp_{\rm I}$
and also $\CKMp_{\rm 2nd}$ when compared with the PDF
errors.\footnote{Note that early LHC measurements of the charge
  asymmetry in SM $W^\pm$ production will very likely reduce the PDF
  errors further.}

We can also understand the general behaviour of $\pRpm$ for different
quark mixing patterns.  As $\wmp$ increases, successively higher
parton momentum fractions $x$ are probed.  While the $u/d$ ratio
increases monotonically with $x$, the $\bar{d}/\bar{u}$ ratio first
increases and then decreases as $x$ increases in the relevant range.
The fractional decrease of the latter overcomes the increase of the
$u/d$ ratio, leading to an overall slight decrease of the parton
luminosity ratio for $\CKMp_{\rm 1st}$, $\CKMp_{\rm I}$ and
$\CKMp_{\rm CKM}$ in the region $\wmp= 3 - 4$~TeV.  On the other hand,
the luminosity ratios for the second and third generation partons are
much flatter functions of $x$, leading to an increasing $\pRpm$
primarily controlled by the $u/d$ ratio for $\CKMp_{\rm II}$, and a
$\pRpm$ ratio close to 1 for $\CKMp_{\rm 2nd}$.

In the next section, we will perform event simulations to quantify the
correlations between $\pRpm$ and $\Rpm$.  We comment at this point
that both values depend on the PDF sets used.  For example, our
simulation is performed using \herwigv~\cite{Bahr:2008pv}, interfaced
with the LHAPDF library \cite{Whalley:2005nh}, with MRST LO** as the
default PDF set.  Compared to MRST LO**, we find significant
differences for the predicted values of $\pRpm$ and hence $\Rpm$ when
using MSTW08 NLO PDFs (interfaced via the LHAPDF library
\cite{Whalley:2005nh}).  This can be understood by considering the
differences in individual PDFS between the recent MSTW08 and earlier
MRST LO** sets.  Of course a detailed simulation of $\Wprime$
production would require a proper NLO calculation using an up-to-date
NLO PDF set such as MSTW08.  We expect that the bulk of the NLO
corrections comes from the differences between the LO and NLO PDF sets
(see, for example, Ref.~\cite{Kom:2010mv}), and so for this reason we
will use MSTW08 NLO as our preferred PDF choice.  Using other PDF
sets, for example CTEQ6.6(NLO)~\cite{Nadolsky:2008zw}, will also lead
to slightly different central values and PDF uncertainties.  The
discrepancies between different PDF sets will hopefully be resolved in
the future.


\section{Numerical study}\label{sec:numerical}

In this section we discuss the basic properties of the signal and
backgrounds calculated using {\herwigv}.  A set of cuts is then
proposed to extract the signal from the background, before the
prospect of quantitatively measuring $\Rpm$ is discussed.

As discussed in the Introduction, the signal is same-sign di-leptons
in association with 2 hadronic jets.  We assume that the heavy
Majorana neutrinos are lighter than the $\Wprime$.  Clearly, an
accurate measurement of $\Rpm$ will require much higher statistics
than a $\Wprime$ discovery.  We will therefore investigate a low mass
range $\wmp=1-2$~TeV.  For simplicity, the masses of the 3 heavy
Majorana neutrinos are assumed to be degenerate, and are given by
$\nm=\frac{1}{2}\wmp$.

To identify a signal event, the basic strategy is to look for a high
$\pt$ SSDL pair and a pair of high $\pt$ jets coming from $ \Wprime
\to l \Maj \to lljj$.  The invariant mass of these four objects will
lead to a resonance peak for the $\Wprime$.  One of the two
lepton-jet-jet combinations will also give a resonance peak for
$\Maj$.

The dominant background is expected to come from $\ttbar$.  Compared
with the studies in \cite{Aad:2009wy,Ferrari:2000sp}, the requirement
of SSDL efficiently suppresses this background.  This is because at
the hadron level, one of the leptons must come from $B$ decay (the
leptons from the two $W$ decays naturally have opposite sign), which
in general does not satisfy a lepton isolation requirement.  Other
potential backgrounds are from $\Wbbj$ and $\WZjj$ production.  The
former process will again be suppressed by a lepton isolation
requirement.  However its production cross section is relatively
large, and therefore it demands more attention.  The latter $\WZjj$
process can lead to SSDL~+~2~jet events if the wrong-sign lepton from
$\Zgam\to l^+l^-$ is not identified, for example when it falls outside
the central tracking region.

Note that since the $\ttbar$ background is produced predominantly by
$gg$ scattering, we have $\Rpm\sim 1$ for this process, i.e. any final
state involving charged leptons is symmetric under change of sign,
whereas $\Rpm$ from $\Wbbj$ and $\WZjj$ is generally $> 1$ since these
arise from $q\bar{q}'$ and $qg$ scatterings.

In terms of other possible backgrounds, we expect that by requiring
SSDL, the $\Zgam$ background discussed in
\cite{Aad:2009wy,Ferrari:2000sp} will be suppressed to a negligible
level.  The $\Wpm\Wpm jj$ background will also be insignificant due to
its small production rate \cite{Gaunt:2010pi,Melia:2010bm}.  Therefore
these processes will not be discussed any further.  We also do not
attempt to study detector effects.  For this reason, the QCD multijet
background is not considered.  A discussion of this background and its
relevance for $\Wprime$ discovery, including a detailed detector
simulation, can be found in Ref.~\cite{Aad:2009wy}.  Note that any
contamination of the $\Wprime \to lljj$ signal from QCD multijet
production will tend to reduce $\Rpm$ since this background is
expected to be charge symmetric.

To simulate the $\Wprime$ signal, we implement the $\Wprime$ model
into \herwig.  The latter is also used to generate both the signal and
the $\ttbar$ background.  We use MSTW08 NLO PDFs together with default
values for other parameters.  The $\WZjj$ and $\Wbbj$ processes are
simulated using \alpgen\,\cite{Mangano:2002ea}, with
\fherwig\,\cite{Corcella:2002jc} then generating the parton showers.
Parton level cuts appropriate to the experimental cuts discussed below
are imposed to speed up the event generation, and we use `out of the
box' SM parameters for these processes.

An event is selected for further analysis based on the following
criteria:
\begin{itemize}
\item the leptons (jets) should lie within pseudorapitidy range
  $|\eta|<2.5\ (4.5)$;
\item a lepton is isolated if $E_{\textrm{{\tiny ISO}}}^l\le
  E_{\textrm{{\tiny ISO}}}^{\textrm{{\tiny min}}}=5$~GeV, where
  $E_{\textrm{{\tiny ISO}}}^l$ is the transverse energy in a cone of
  $R=0.4$ surrounding the lepton;
\item the 2 hightest $\pt$ leptons must have the same sign and
  flavour;
\item the 2 highest $\pt$ jets and the 2 highest $\pt$ leptons must be
  separated by $\Delta R \ge 0.4$.
\end{itemize}

In the above, an isolated lepton is assumed to be identified with
100\% efficiency when $\ptl > 10$~GeV.  Muons are assumed to be
invisible when $|\eta_\mu| > 2.5$.  Our jet reconstructions are
performed using \fastjet\,\cite{FASTJET} using the anti-$k_T$
algorithm \cite{Cacciari:2008gp} with $R=0.4$.

We then apply further cuts to improve the signal/background ratio:
\begin{enumerate}
\item select events with $\ge$ 2 isolated leptons with $\ptl > 75$~GeV
  and $\ge$ 2 jets with $\ptj > 50$~GeV;
\item veto an event if a third lepton with $\ptl > 10$~GeV is present
  which has the same flavour but opposite charge to the two hardest
  $\ptl$ leptons;
\item retain an event when the invariant mass of the 2 leptons and 2
  jets, $\mlljj$, satisfies \bea 0.7\mwrec < &\mlljj& < 1.2\mwrec.
  \nonumber \eea Here $\mwrec$ is the value of the reconstructed
  $\Wprime$ mass.
\end{enumerate}
Of course the backgrounds considered will have missing transverse
energy $\MET$ already at the parton level.  Furthermore leptons from
$B$ decays can have displaced vertices.  These can be used as
additional handles to further suppress the backgrounds.  However we
refrain from doing so here in an attempt to remain conservative.  The
leading-order multiparticle final-state backgrounds are known to
suffer from scale uncertainties, and so the values quoted below should
be treated as rough estimates only.

Due to tight lepton isolation and hard lepton $\pt$ cut, we find that
the $\Wbbj$ background is suppressed to negligible level.  We provide
a numerically more conservative estimate using \mcfm\,\cite{MCFM} (at
leading order).  This parton level simulation does not model
semi-leptonic decay of the $b$s, rather a `rule of thumb' efficiency
factor $\epsb\sim 1/200$ is used for the probability of obtaining an
isolated lepton of a given flavour (i.e. $e$ or $\mu$) from $B$ decay
suggested recently in \cite{Sullivan:2010jk}.

In Table~\ref{tab:event_cut} we display the effect of the selection
cuts for $\wmp=1.0, 1.5, 2.0$~TeV with the quark mixing structures
$\CKMp_{\rm I}$ and $\CKMp_{\rm II}$.  $\sigma_{1,2,3}$ refer to the
cross sections after applying the cuts $1$ (hard leptons and jets),
$2$ (opposite-sign, same-flavour veto (OSSFV)) and $3$ ($lljj$
invariant mass constraint) defined above, including the $++$ and $--$
contributions from both electrons and muons.  In general, the signal
cross sections for $\CKMp_{\rm II}$ are smaller than those for
$\CKMp_{\rm I}$, reflecting the different PDF combinations entering
the processes via off-diagonal and diagonal flavour couplings
respectively.  Applying the $\wmp$ invariant mass cut further
suppresses the backgrounds by a factor of a few, while retaining most
of the signal as expected.  Our more conservative estimate of the
$\Wbbj$ background is still relatively small compared with the
$\ttbar$ background.  For the $\WZjj$ background, the OSSFV is shown
to be efficient.  In the following we will therefore neglect the
$\Wbbj$ and $\WZjj$ contributions when performing the $\Rpm$
estimation.

In Fig.~\ref{fig:wprimeDist} we show some sample kinematic
distributions (normalised to 5~$\ifb$) from different processes.  For
clarity, only $\Wprime$ signals with $\CKMp_{\rm I}$ mixing and the
dominant $\ttbar$ background are shown.  We see clearly the $\Wprime$
and $\Maj$ resonance peaks from the invariant masses of $lljj$ and
$ljj$ respectively.  The $ljj$ peak from the second hardest lepton is
more pronounced due to our assumed relation $\nm=\frac{1}{2}\wmp$,
which results in higher average $\pt$ for the lepton from the decay of
the $\Wprime$ than that from the decay of the $\Maj$.

\begin{table}
  \centering
  \renewcommand{\arraystretch}{1.2}
  \begin{tabular}{|c|c|cccc|}
    \hline
    $\wmp$ & Process & $\sigma_{\rm tot}$ & $\sigma_{1}$ & $\sigma_{2}$ & $\sigma_{3}$\\
    \hline\hline
    \multirow{5}{*}{1.0 TeV}&$\Wprime(\CKMp_{\rm I})$  & 4.78$\cdot 10^3$  & 1.19$\cdot 10^3$   & 1.15$\cdot 10^3$ & 1.02$\cdot 10^3$\\
    &$\Wprime(\CKMp_{\rm II})$ & 2.62$\cdot 10^3$  & 647 & 622 & 542\\
    \cline{2-6}
    &$\ttbar$                & 6.06$\cdot 10^5$   & 11 & 9.7 & 2.8\\
    &$\WZjj$                 & -     & 3.0 & 0.83 & 0.37\\
    &$\Wbbj$                 & -     & 115$\epsb$ & 115$\epsb$ & 80$\epsb$\\
    \hline\hline
    \multirow{5}{*}{1.5 TeV}& $\Wprime(\CKMp_{\rm I})$  & 882 & 289  & 276 & 253\\
    &$\Wprime(\CKMp_{\rm II})$ & 411 & 135  & 128 & 116\\
    \cline{2-6}
    &$\ttbar$                & 6.06$\cdot 10^5$   & 11  & 9.7 & 1.4\\
    &$\WZjj$                 & -     & 3.0 & 0.83 & 0.22\\
    &$\Wbbj$                 & -     & 115$\epsb$ & 115$\epsb$ & 41$\epsb$\\
    \hline\hline
    \multirow{5}{*}{2.0 TeV}&$\Wprime(\CKMp_{\rm I})$  & 226 & 81.8 & 77.8 & 72.9\\
    &$\Wprime(\CKMp_{\rm II})$ & 92.8& 33.6 & 31.7 & 29.3\\
    \cline{2-6}
    &$\ttbar$                & 6.06$\cdot 10^5$   & 11  & 9.7 & 0.53\\
    &$\WZjj$                 & -     & 3.0 & 0.83 & 0.14\\
    &$\Wbbj$                 & -     & 115$\epsb$ & 115$\epsb$ & 19$\epsb$\\
    \hline
  \end{tabular}
  \caption{Cross sections (in $\fb$) for different signal and background  processes at 14~TeV
    LHC.  All 4 same-sign, same-flavour di-lepton channels are
    included.  $\sigma_{1,2,3}$ correspond to the cross sections after
    applying cuts $1,2,3$ described in the text.  $\epsb$ denotes the
    efficiency in extracting an isolated lepton from a $b$ quark and
    is estimated to be $\sim 1/200$ \cite{Sullivan:2010jk}.  See
    Table~\protect\ref{tab:RvsR} for the simulated $\Rpm$ values.}
  \label{tab:event_cut}
\end{table}

\begin{figure}[!htp]
  \begin{center}
      \begin{tabular}{cc}
        \subfigure[$m_{l_1jj}$]{
          \scalebox{0.76}{
            \includegraphics{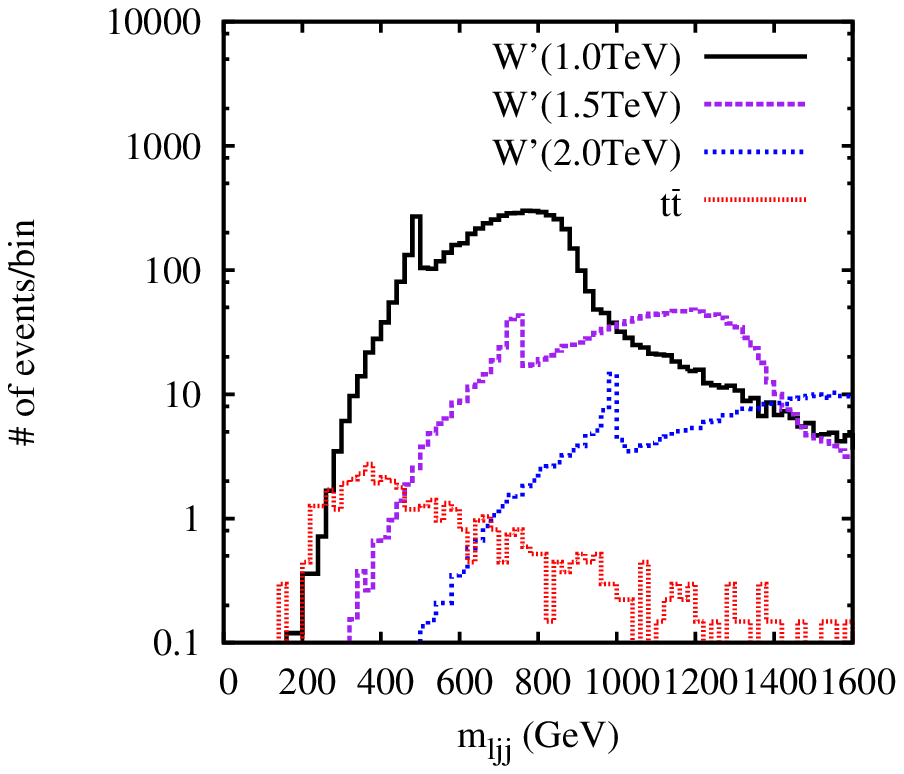}
          }
        }
        &
        \subfigure[$m_{l_2jj}$]{
          \scalebox{0.76}{
            \includegraphics{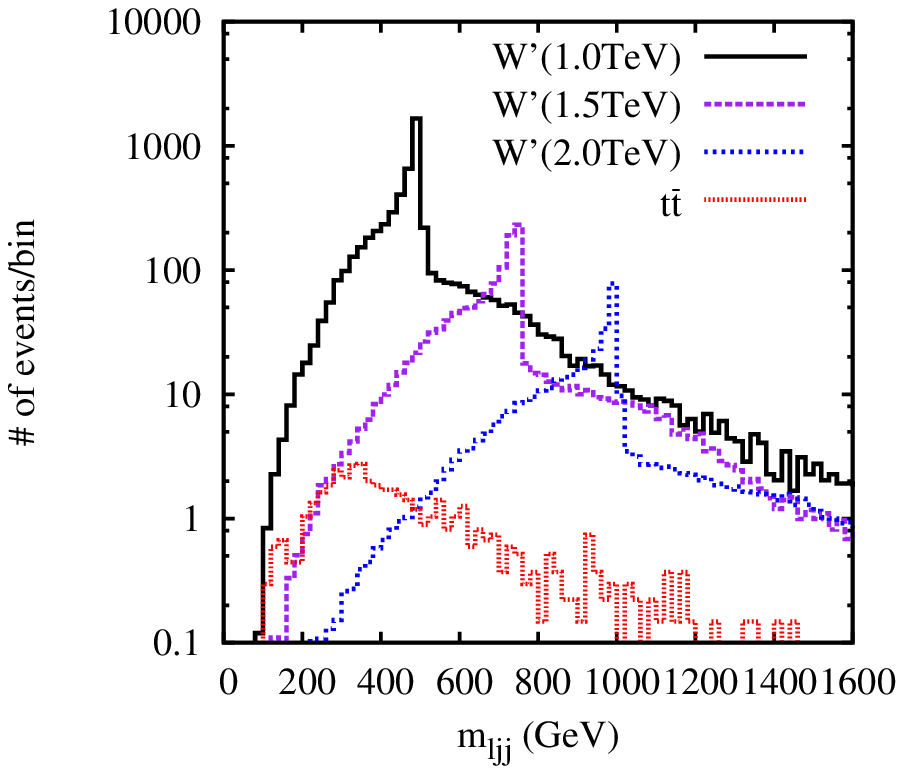}
          }
        }
        \\
        \subfigure[$m_{lljj}$]{ 
          \scalebox{0.76}{
            \includegraphics{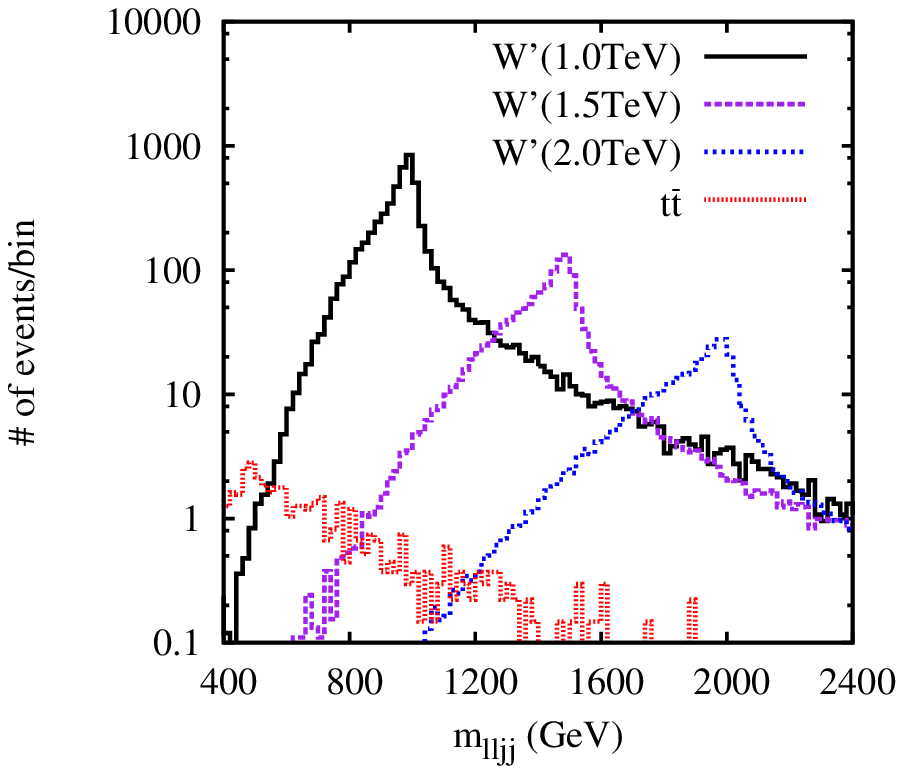} }
        }
        &
        \subfigure[$m_{ll}$]{
          \scalebox{0.76}{
            \includegraphics{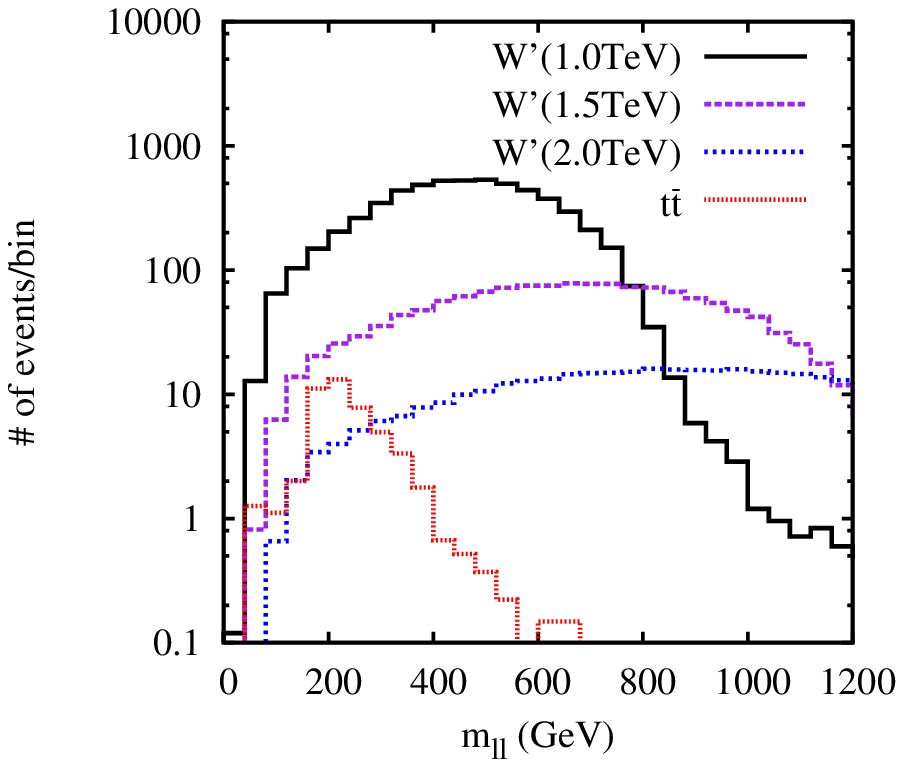}
          }
        }
        \\
        \subfigure[$\pt$ (2nd hardest lepton)]{
          \scalebox{0.76}{
            \includegraphics{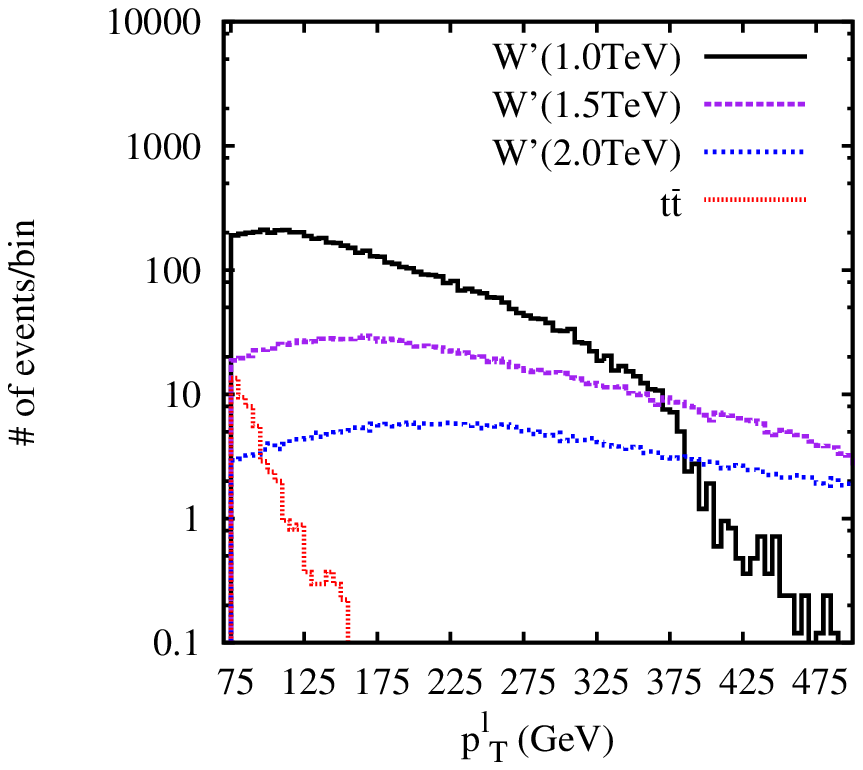}
          }
        }
        &
        \subfigure[$\pt$ (2nd hardest jet)]{
          \scalebox{0.76}{
            \includegraphics{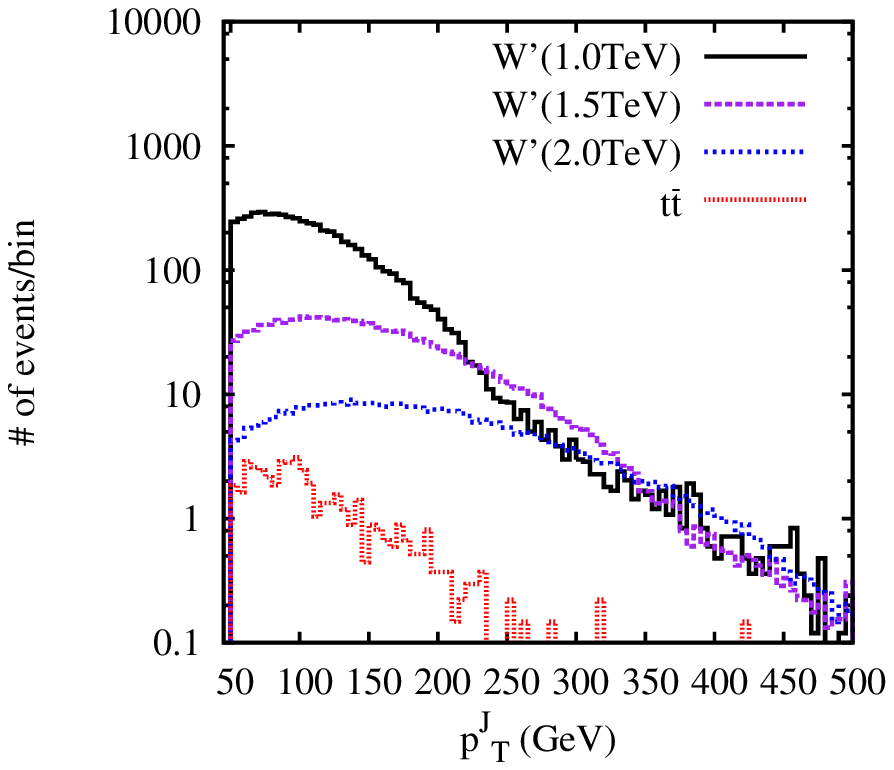}
          }
        }
        \\
      \end{tabular}
      \caption{Selected kinematic distributions for the same-sign
        dilepton events.  The number of events are normalised to
        $5\,\ifb$.}
      \label{fig:wprimeDist}
  \end{center}
\end{figure}

We also see that more pure samples can readily be obtained by varying the cut values,
for example the $\pt$ of the second hardest lepton, as a function of
$\wmp$.  Additional cuts, for example on the invariant mass of $\nm$,
may also be useful.  Given that we already have fairly pure samples we
refrain from optimising the cuts further.  Other strategies for
extracting the signal can be found in Refs.~\cite{Aad:2009wy,Ferrari:2000sp}.

As discussed earlier, the focus of this work is on the extraction of
$\Rpm$.  In Table~\ref{tab:RvsR}, a comparison between $\pRpm$ and the
simulated $\Rpm$ is shown.  The cross-section ratios at the hard
process level obtained from \herwigv\, are well within the PDF
uncertainties of the parton luminosity ratio predictions, while the
values of $\Rpm$ are systematically smaller than that of $\pRpm$.  The
latter is due to the finite acceptance in lepton pseudorapidity and
the fact that the average parton $x$ for an up quark is higher than
that of a down quark.  The $\Wprime^+$, and hence the $l^+$, will tend
to have a broader rapidity distribution than that of the $\Wprime^-$
and $l^-$, leading to a larger acceptance loss of the $l^+$.  It is
expected that while the extracted $\Rpm$ values will depend slightly
on the cuts applied, they will nevertheless track $\pRpm$ closely.

\begin{table}
  \centering
  \renewcommand{\arraystretch}{1.2}
  \begin{tabular}{|c|cc|cc||cc|}
    \hline
    \multicolumn{1}{|c|}{}&\multicolumn{2}{c|}{$\CKMp_{\rm I}$}&\multicolumn{2}{c||}{$\CKMp_{\rm II}$}&$\ttbar$&$\WZjj$\\
    \cline{2-7}
    $\wmp$ 
    & $\pRpm$ & $\Rpm$ 
    & $\pRpm$ & $\Rpm$ 
    & $\Rpm$ & $\Rpm$ \\
    \hline\hline
    \renewcommand{\arraystretch}{1.0}
    1.0 TeV & 2.12(4) & 1.99(1) & 1.92(4) & 1.79(2) &1.0(1)&1.2(2)\\
    1.5 TeV & 2.50(6) & 2.42(3) & 2.21(7) & 2.13(4) &1.0(2)&1.1(3)\\
    2.0 TeV & 2.82(9) & 2.74(7) & 2.49(10)& 2.40(10)&1.1(4)&1.2(3)\\
    \hline
  \end{tabular}
  \caption{Comparison between parton luminosity ratio $\pRpm$ and
    observed charged asymmetry ratios $\Rpm$ (MSTW NLO) at 14 TeV LHC.
    For the signal, Poisson uncertainties for the $\Rpm$'s at
    100 $\ifb$ and 1 $\sigma$ PDF uncertainties for $\pRpm$ are
    displayed in brackets.  For the backgrounds, we show the
    statistical uncertainties based on the number of events generated.
    The $\Wbbj$ results are not shown due to too few statistics.}
  \label{tab:RvsR}
\end{table}

In Fig.~\ref{fig:RvsM}, we compare $\pRpm$ and $\Rpm$ as a function of
$\wmp$, with quark mixings $\CKMp_{\rm I}$ and $\CKMp_{\rm II}$ as
examples, at 14~TeV LHC with an assumed luminosity of 30~$\ifb$.  For
simplicity, we have not included the background contributions to
$\Rpm$.  From Table~\ref{tab:RvsR} we see that the latter have $\Rpm$
values much closer to 1.  Together with results in
Table~\ref{tab:event_cut}, including background contributions would
lower the $\Rpm$ values slightly by $\ord (1 - 2)$\%.  We see that for
$\wmp$ close to 1~TeV, the two mixing structures are clearly
distinguishable.  At this luminosity, the uncertainties on $\Rpm$ may
be expected to come primarily from the PDFs.  Statistical
uncertainties become comparable at $\wmp\sim 1.5$~TeV.  However at
this luminosity, the ability to distinguish $\CKMp_{\rm I}$ from
$\CKMp_{\rm II}$ becomes limited when $\wmp$ approaches 2~TeV.

\begin{figure*}[!ht]
  \begin{center}
    \scalebox{1.2}{ \includegraphics{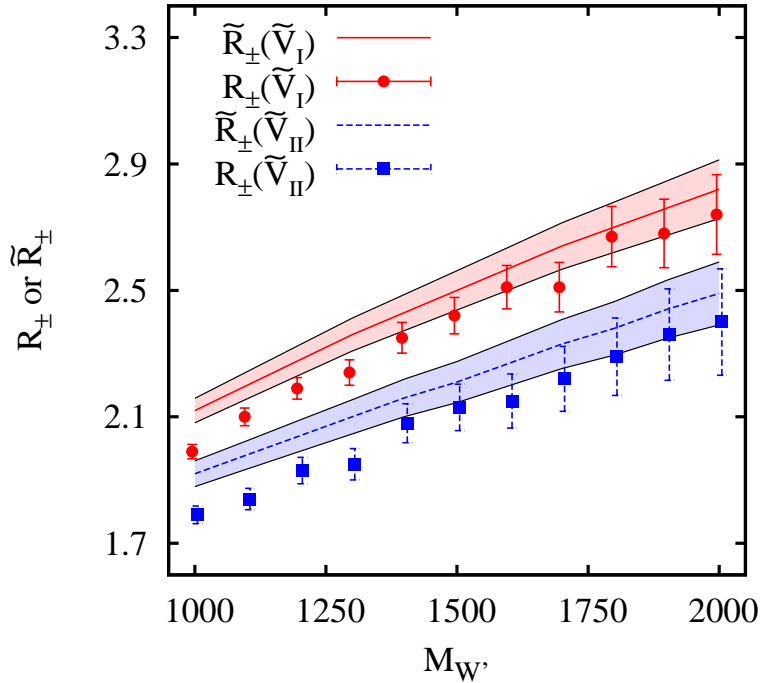} }
    \caption{Comparison between parton luminosity ratio $\pRpm$ and
      observed charged asymmetry ratios $\Rpm$ (MSTW NLO) at 14 TeV
      LHC as a function of $\wmp$.  Results from two quark mixing
      patterns, $\CKMp_{\rm I}$ (upper, red curves) and $\CKMp_{\rm
        II}$ (lower, green curves) are shown.  The colour bands
      correspond to 1 $\sigma$ PDF uncertainties, while the error bars
      are Poisson uncertainties on $\Rpm$ at 30 $\ifb$.}
    \label{fig:RvsM} 
  \end{center}
\end{figure*}

Finally, given the fact that in the $\wmp=\ord (1.0)$~TeV region the
$\Wprime$ can be easily detected, we discuss the extent to which early
LHC data (i.e. $\sim 1\,\ifb$ at 7~TeV) can measure $\Rpm$.  Measuring
$\Rpm$ at different centre of mass energies will probe different
parton $x$ values, and hence could provide additional information to
discriminate different PDF choices.

\begin{table}
  \centering
  \renewcommand{\arraystretch}{1.2}
  \begin{tabular}{|c|c|cc|cc|}
    \hline
    $\wmp$ & Process & $\sigma_{\rm tot}$ & $\sigma_{3}$ & $\pRpm$ & $\Rpm$ \\
    \hline\hline
    \multirow{2}{*}{0.8 TeV}&$\Wprime(\CKMp_{\rm I})$  & 2.88$\cdot 10^3$  & 5.1$\cdot 10^2$ & 2.57(7) & 2.4(2)\\
                            &$\Wprime(\CKMp_{\rm II})$ & 1.26$\cdot 10^3$  & 2.2$\cdot 10^2$ & 2.26(7) & 2.2(3)\\
    \hline\hline
    \multirow{2}{*}{1.0 TeV}&$\Wprime(\CKMp_{\rm I})$  & 9.98$\cdot 10^2$  & 2.4$\cdot 10^2$ & 2.81(9) & 2.8(4)\\
                            &$\Wprime(\CKMp_{\rm II})$ & 4.0$\cdot 10^2$   & 92              & 2.49(10) & 2.4(5)\\
    \hline\hline
    \multirow{2}{*}{1.2 TeV}&$\Wprime(\CKMp_{\rm I})$  & 381               & 105             & 2.99(12) & 2.9(6)\\
                            &$\Wprime(\CKMp_{\rm II})$ & 142               & 38              & 2.71(14) & 2.6(10)\\
    \hline
  \end{tabular}
  \caption{Cross sections (in $\fb$), $\pRpm$ and $\Rpm$ for different
    $\Wprime$ models at 7~TeV LHC.  All 4 same-sign, same-flavour
    di-lepton channels are included.  $\sigma_{3}$ corresponds to the
    cross section after applying cut $3$ described in the text.  The
    brackets for $\pRpm$ are $\pm 1\, \sigma$ PDF uncertainties, whereas
    those for $\Rpm$ are Poisson uncertainties at $1\,\ifb$
    luminosity.  After applying cut 2, the background processes are
    suppressed below 1~$\fb$, and hence their values are not shown.}
  \label{tab:event_cut_7TeV}
\end{table}

In Table~\ref{tab:event_cut_7TeV}, we show the results for the signal
processes at 7~TeV, assuming an integrated luminosity of 1~$\ifb$.
Using the same cuts as the 14~TeV case, the backgrounds we have
considered are again shown to be tiny ($\sigma(\ttbar)<1\,\fb$,
$\sigma(\WZjj)<0.2\,\fb$ and $\sigma(\Wbbj)$ negligible after cut 2).
We see that while $\Wprime$ could be discovered easily, the assumed
total luminosity is only able to provide marginal discriminating power
between $\CKMp_{\rm I}$ and $\CKMp_{\rm II}$ for $\wmp$ below 1~TeV.
However such low values of $\wmp$ may have difficulty evading the
bounds from EW precision observables already discussed.


\section{Conclusions}

In this paper we have demonstrated how by measuring the charge
asymmetry ratio $\Rpm$ at the LHC, non-trivial information on the
couplings of new heavy charged resonances to quarks may be obtained.
We have focussed in particular on the $\Wprime$ model, confirming
expectations in the literature that in the $\wmp\sim\ord ({\rm TeV})$
region the physics backgrounds can be highly suppressed by applying
appropriate cuts in the same-sign di-lepton + 2 jet channel.  As a
result, the measured $\Rpm$ value and the parton luminosity ratio
$\pRpm$ are related in a directly correlated manner.  Since the value
of $\pRpm$ can be calculated precisely given a particular flavour
mixing structure, we have established a method to constrain the quark
mixing given a measured value of $\Rpm$. Although we have considered
overall event rates only, with higher luminosity it should be possible
to study the rapidity dependence of the charge asymmetry ratio and
this would provide additional constraints.

Quantitatively, we have shown that at 14~TeV LHC three classes of
representative flavour structures, namely $\CKMp_{({\rm I,CKM,1st})},
\CKMp_{\rm II}$ and $\CKMp_{\rm 2nd}$, can be clearly distinguished
with realistic luminosities.  However the prospect at 7~TeV LHC is
less optimistic due to the lower overall luminosity as well as the
larger PDF uncertainties in the higher parton $x$ region being probed.

A knowledge of quark flavour mixing structure will clearly have
important implications on EW precision observables.  Interestingly, if
$\Wprime$ is relevant for neutrinoless double beta decay ($\nubb$),
then measurement of $\Rpm$\footnote{In this case, one should focus on
  the di-electron channel.} will be crucial in estimating its
contribution to $\nubb$, as only the first generation quarks
participate in this low energy process.  The same consideration will
apply in relating an observation of resonant selectron production in
R-parity violating SUSY models to its contribution to $\nubb$
\cite{LHC_nubb}.

Finally, the measurement of $\Rpm$ can be extended to signals other
than resonance production.  We will study the application of this
observable in other NP models, for example SUSY, in forthcoming
publications.


\section*{Acknowledgements}

This work has been partially supported by the Isaac Newton Trust at
the University of Cambridge.  We thank the members of the Cambridge
SUSY working group for useful conversations.  CHK thanks Andreas
Papaefstathiou for invaluable help in using \herwig.



\begin{thebibliography}{99}

\bibitem{Kom:2010mv}
  C.~H.~Kom and W.~J.~Stirling,
  Eur.\ Phys.\ J.\  C {\bf 69} (2010) 67
  [arXiv:1004.3404 [hep-ph]].

\bibitem{Martin:2009iq}
  A.~D.~Martin, W.~J.~Stirling, R.~S.~Thorne and G.~Watt,
  Eur.\ Phys.\ J.\  C {\bf 63} (2009) 189
  [arXiv:0901.0002 [hep-ph]].

\bibitem{MCFM}
  J.~M.~Campbell and R.~K.~Ellis,
  Phys.\ Rev.\  D {\bf 65} (2002) 113007
  [arXiv:hep-ph/0202176];
\\
  J.~M.~Campbell and R.~K.~Ellis,
  http://mcfm.fnal.gov/.

\bibitem{WnJ_NLODixon}
  C.~F.~Berger {\it et al.},
  Phys.\ Rev.\ Lett.\  {\bf 102} (2009) 222001
  [arXiv:0902.2760 [hep-ph]];
\\
  C.~F.~Berger {\it et al.},
  Phys.\ Rev.\  D {\bf 80} (2009) 074036
  [arXiv:0907.1984 [hep-ph]].
\\
  C.~F.~Berger {\it et al.},
  arXiv:1009.2338 [hep-ph].

\bibitem{Altarelli:1989ff}
  G.~Altarelli, B.~Mele and M.~Ruiz-Altaba,
  Z.\ Phys.\  C {\bf 45} (1989) 109
  [Erratum-ibid.\  C {\bf 47}, 676 (1990)].

\bibitem{Ramond:1983qu}
  P.~Ramond,
  Ann.\ Rev.\ Nucl.\ Part.\ Sci.\  {\bf 33} (1983) 31.

\bibitem{Dreiner:2000vf}
  H.~K.~Dreiner, P.~Richardson and M.~H.~Seymour,
  Phys.\ Rev.\  D {\bf 63}, 055008 (2001)
  [arXiv:hep-ph/0007228].

\bibitem{LRSM}
  J.~C.~Pati and A.~Salam,
  Phys.\ Rev.\  D {\bf 10}, 275 (1974)
  [Erratum-ibid.\  D {\bf 11}, 703 (1975)].
\\
  R.~N.~Mohapatra and J.~C.~Pati,
  Phys.\ Rev.\  D {\bf 11}, 566 (1975).
\\
  R.~N.~Mohapatra and J.~C.~Pati,
  Phys.\ Rev.\  D {\bf 11}, 2558 (1975).
\\
  G.~Senjanovic and R.~N.~Mohapatra,
  Phys.\ Rev.\  D {\bf 12}, 1502 (1975).
\\
  G.~Senjanovic,
  Nucl.\ Phys.\  B {\bf 153}, 334 (1979).

\bibitem{CDF_Wprime}
  T.~Aaltonen {\it et al.}  [The CDF Collaboration],
  Phys.\ Rev.\ Lett.\  {\bf 104} (2010) 241801
  [arXiv:1004.4946 [hep-ex]].
\\
  T.~Aaltonen {\it et al.}  [CDF Collaboration],
  Phys.\ Rev.\ Lett.\  {\bf 103} (2009) 041801
  [arXiv:0902.3276 [hep-ex]].
\\
  A.~Abulencia {\it et al.}  [CDF Collaboration],
  Phys.\ Rev.\  D {\bf 75} (2007) 091101
  [arXiv:hep-ex/0611022].

\bibitem{D0_Wprime}
  V.~M.~Abazov {\it et al.}  [D0 Collaboration],
  Phys.\ Rev.\ Lett.\  {\bf 100} (2008) 031804
  [arXiv:0710.2966 [hep-ex]].
\\
  V.~M.~Abazov {\it et al.}  [D0 Collaboration],
  Phys.\ Rev.\ Lett.\  {\bf 100} (2008) 211803
  [arXiv:0803.3256 [hep-ex]].

\bibitem{Aad:2009wy}
  G.~Aad {\it et al.}  [The ATLAS Collaboration],
  arXiv:0901.0512 [hep-ex].

\bibitem{Ferrari:2000sp}
  A.~Ferrari {\it et al.},
  Phys.\ Rev.\  D {\bf 62} (2000) 013001.

\bibitem{Amsler:2008zzb}
  C.~Amsler {\it et al.}  [Particle Data Group],
  Phys.\ Lett.\  B {\bf 667} (2008) 1.

\bibitem{Beall:1981ze}
  G.~Beall, M.~Bander and A.~Soni,
  Phys.\ Rev.\ Lett.\  {\bf 48} (1982) 848.

\bibitem{Papaefstathiou:2009sr}
  A.~Papaefstathiou and O.~Latunde-Dada,
  JHEP {\bf 0907} (2009) 044
  [arXiv:0901.3685 [hep-ph]].

\bibitem{Maiezza:2010ic}
  A.~Maiezza, M.~Nemevsek, F.~Nesti and G.~Senjanovic,
  arXiv:1005.5160 [hep-ph].

\bibitem{Langacker:1989xa}
  P.~Langacker and S.~Uma Sankar,
  Phys.\ Rev.\  D {\bf 40} (1989) 1569.

\bibitem{Buras:2010pz}
  A.~J.~Buras, K.~Gemmler and G.~Isidori,
  arXiv:1007.1993 [hep-ph].

\bibitem{Aaltonen:2007he}
  T.~Aaltonen {\it et al.}  [CDF Collaboration],
  Phys.\ Rev.\ Lett.\  {\bf 100} (2008) 161802
  [arXiv:0712.2397 [hep-ex]].
\\
  V.~M.~Abazov {\it et al.}  [D0 Collaboration],
  Phys.\ Rev.\ Lett.\  {\bf 101} (2008) 241801
  [arXiv:0802.2255 [hep-ex]].
\\
  V.~M.~Abazov {\it et al.}  [D0 Collaboration],
  Phys.\ Rev.\  D {\bf 82} (2010) 032001
  [arXiv:1005.2757 [hep-ex]].

\bibitem{Bahr:2008pv}
  M.~Bahr {\it et al.},
  Eur.\ Phys.\ J.\  C {\bf 58} (2008) 639
  [arXiv:0803.0883 [hep-ph]].

\bibitem{Whalley:2005nh}
  M.~R.~Whalley, D.~Bourilkov and R.~C.~Group,
  arXiv:hep-ph/0508110.

\bibitem{Nadolsky:2008zw}
  P.~M.~Nadolsky {\it et al.},
  Phys.\ Rev.\  D {\bf 78} (2008) 013004
  [arXiv:0802.0007 [hep-ph]].

\bibitem{Gaunt:2010pi}
  J.~R.~Gaunt, C.~H.~Kom, A.~Kulesza and W.~J.~Stirling,
  Eur.\ Phys.\ J.\  C {\bf 69} (2010) 53
  [arXiv:1003.3953 [hep-ph]].

\bibitem{Melia:2010bm}
  T.~Melia, K.~Melnikov, R.~Rontsch and G.~Zanderighi,
  arXiv:1007.5313 [hep-ph].

\bibitem{Mangano:2002ea}
  M.~L.~Mangano, M.~Moretti, F.~Piccinini, R.~Pittau and A.~D.~Polosa,
  JHEP {\bf 0307} (2003) 001
  [arXiv:hep-ph/0206293].

\bibitem{Corcella:2002jc}
  G.~Corcella {\it et al.},
  arXiv:hep-ph/0210213.

\bibitem{FASTJET}
  M.~Cacciari and G.~P.~Salam
  Phys.\ Lett.\ B  {\bf 641} (2006) 57.
  [arXiv:hep-ph/0512210]
  http://fastjet.fr/

\bibitem{Cacciari:2008gp}
  M.~Cacciari, G.~P.~Salam and G.~Soyez,
  JHEP {\bf 0804} (2008) 063
  [arXiv:0802.1189 [hep-ph]].

\bibitem{Sullivan:2010jk}
  Z.~Sullivan and E.~L.~Berger,
  Phys.\ Rev.\  D {\bf 82} (2010) 014001
  [arXiv:1003.4997 [hep-ph]].

\bibitem{LHC_nubb}
  B.~C.~Allanach, C.~H.~Kom and H.~P\"as,
  Phys.\ Rev.\ Lett.\  {\bf 103} (2009) 091801
  [arXiv:0902.4697 [hep-ph]].
\\
  B.~C.~Allanach, C.~H.~Kom and H.~P\"as,
  JHEP {\bf 0910} (2009) 026
  [arXiv:0903.0347 [hep-ph]].

\end{thebibliography}
\end{document}
